\documentclass[12pt,a4paper]{article} 

\usepackage{graphicx}

\usepackage{epsfig}
\usepackage{subfig} 
\usepackage{cite}

\usepackage{color} 
\usepackage[dvipsnames]{xcolor}

\usepackage{amssymb}
\usepackage{amsmath}
\usepackage{authblk}
\usepackage{doi}
\usepackage{dsfont}
\usepackage{setspace}

\usepackage{verbatim}

 \usepackage{slashed}
\usepackage[normalem]{ulem} 

\newcommand{\mr}{\mathrm}
\newcommand{\mc}{\mathcal}
\newcommand{\pol}{\mr{pol}}
\newcommand{\bbar}{\overline{B}}
\newcommand{\psn}{\mathbf{\Phi}_n}

\newcommand{\psrad}{\mathbf{\Phi}_{\mathrm{rad}}}

\newcommand{\xif}{\xi_\mr{F}}
\newcommand{\xir}{\xi_\mr{R}}

\newcommand{\MGAMCNLO}{{\tt{MadGraph5\_aMC@NLO}}} 
\newcommand{\MADGRAPH}{{\tt{MadGraph}}}

\newcommand{\POWHEG}{{\tt{POWHEG}}}
\newcommand{\POWHEGBOX}{{\tt{POWHEG~BOX}}}
\newcommand{\PBOX}{\POWHEGBOX}
\newcommand{\POWHEGBOXVV}{{\tt{POWHEG~BOX~V2}}}

\newcommand{\PYTHIA}{{\tt{PYTHIA}}}
\newcommand{\PYTHIAS}{{\tt{PYTHIA6}}}

\newcommand{\LHE}{{\tt{LHE}}}

\newcommand{\beq}{\begin{equation}}
\newcommand{\eeq}{\end{equation}}

\newcommand{\bea}{\begin{eqnarray}}
\newcommand{\eea}{\end{eqnarray}}

\newcommand{\GeV}{\mr{GeV}}

\newcommand{\etaj}{\eta^\mr{jet}}
\newcommand{\etajone}{\eta^\mr{jet\,1}}
\newcommand{\etajtwo}{\eta^\mr{jet\,2}}

\newcommand{\ptjone}{p_T^{\mr{jet\,1}}}
\newcommand{\ptjtwo}{p_T^{\mr{jet\,2}}}
\newcommand{\ptjthree}{p_T^{\mr{jet\,3}}}
\newcommand{\phij}{\Delta\phi_{jj}}

\newcommand{\mjj}{M_{jj}}

\begin{document}

\begin{titlepage}

\vskip 0.8cm

\begin{center}
{\Large \bf A POWHEG generator for di-jet production in \\[2ex]
polarized proton-proton collisions}

\vskip 1.cm

{\large {
{\bf Ignacio Borsa}{\footnote{\tt ignacio.borsa@itp.uni-tuebingen.de}}, 
{\bf David Betz}{\footnote{\tt david.betz@student.uni-tuebingen.de}}, 
{\bf Barbara J\"ager}{\footnote{\tt jaeger@itp.uni-tuebingen.de}}}}

\vskip 1.cm

{\it Institute for Theoretical Physics, University of T\"ubingen,
Auf~der~Morgenstelle~14, 72076~T\"ubingen, Germany 
} 

\vspace{1.cm}

{\bf Abstract}

\end{center}

\vspace{.5cm}
We present a new Monte-Carlo generator for the simulation of di-jet production in polarized proton-proton collisions at the next-to-leading order in QCD matched to parton showers using the framework of the \PBOX{}. With this program we compute a variety of observables of immediate relevance for the spin program of the Relativistic Heavy Ion Collider at Brookhaven National Laboratory. While parton-shower effects are generally small, we find that in some search regions their inclusion improves agreement of predictions with data. Moreover, we provide a critical assessment of selection criteria applied in experiment in the light of perturbative stability. 

\noindent

\end{titlepage}

\newpage

{\small \tableofcontents}

\section{Introduction}
Understanding the spin structure of the proton remains a central goal of contemporary hadronic physics. Within the framework of perturbative QCD, the individual contributions from quarks and gluons to the total spin of the proton are encoded in terms of polarized parton distribution functions (PDFs), which can be extracted in global analyses of polarized data in complete analogy to their unpolarized counterparts \cite{Lampe:1998eu, Aidala:2012mv}. While measurements in longitudinally polarized lepton-hadron scattering, as well as in longitudinally polarized proton-proton collisions, have led to significant progress in the extraction of these distributions, our understanding of how the proton spin emerges from that of its constituents is still incomplete \cite{deFlorian:2014yva, Nocera:2014gqa, Bertone:2024taw, Borsa:2024mss, Cruz-Martinez:2025ahf, Cocuzza:2025qvf}. In particular, sizeable uncertainties persist in the gluon helicity distribution.

Among the observables used to constrain polarized PDFs, high-$p_T$ jet production in longitudinally polarized proton-proton collisions plays a particularly important role. Such processes receive contributions from gluon-initiated subprocesses already at leading order and therefore offer enhanced sensitivity to the gluon distribution compared to inclusive observables measured in lepton-hadron collisions. For this reason, jet observables have been a cornerstone of the spin program at the Relativistic Heavy-Ion Collider (RHIC) \cite{Aschenauer:2015eha}, with measurements of single-jet and di-jet production performed by the PHENIX \cite{PHENIX:2010aru} and STAR \cite{STAR:2006:JetALL:PRL97,STAR:2008:JetALL:PRL100, STAR:2012hth, STAR:2014wox, STAR:2016kpm, STAR:2018yxi, STAR:2019yqm, STAR:2021mfd, STAR:2021mqa, STAR:2025:DijetIntermediateALL:PRD112} collaborations at center-of-mass energies of $\sqrt{s} = 200~\GeV$ and $\sqrt{s} = 510~\GeV$. In particular, di-jet measurements allow for a more differential access to the partonic kinematics than inclusive jets, enabling a separation of the momentum fractions carried by the initial-state partons and thus providing enhanced sensitivity to polarized PDFs \cite{DeFlorian:2019xxt}.

From the theoretical perspective, next-to-leading order (NLO) QCD calculations for polarized jet and di-jet production have been available for some time \cite{deFlorian:1998poljets, deFlorian:1999polNLO, Mukherjee:2003pf,Jager:2004jetpol} and have been successfully employed in phenomenological studies and global PDF analyses \cite{deFlorian:2014yva, Nocera:2014gqa, DeFlorian:2019xxt, Bertone:2024taw, Borsa:2024mss, Cruz-Martinez:2025ahf, Cocuzza:2025qvf}. This stands in contrast to the unpolarized case, for which not only NLO predictions for up to three-jet production exist \cite{Ellis:1990jetNLO, Ellis:1992dijetNLO, Nagy:2003threejet}, but also NNLO-accurate calculations are available for single-jet and di-jet observables \cite{Currie:2017dijetNNLO, GehrmannDeRidder:2019triplediff}.

While fixed-order predictions provide reliable results for sufficiently inclusive observables, they can become unstable in regions of phase space sensitive to soft or collinear radiation. Moreover, experimental analyses typically rely on fully exclusive event simulations, including parton-shower effects, hadronization, and detector corrections, which cannot be consistently addressed within a purely fixed-order framework. The matching of NLO calculations with parton showers offers a way to overcome these limitations by combining the accuracy of fixed-order perturbation theory with an all-order resummation of logarithmically enhanced contributions from multiple emissions \cite{Frixione:2002ik, Nason:2004rx}. The  implementation in a multi-purpose Monte-Carlo generator further allows to simulate events at the hadron-level and potentially consider detector effects.
The \POWHEG{} method \cite{Frixione:2007vw, Alioli:2010xd} provides a well-established approach to achieve such matching, and has been extensively applied to a wide range of observables in unpolarized proton-proton collisions, including dijet production \cite{Alioli:2010xa}.

Similar tools are largely absent for longitudinally polarized processes in hadronic collisions, with simulations used in experimental studies often relying on unpolarized parton-shower generators supplemented by reweighting procedures based on leading-order spin asymmetries \cite{STAR:2025:DijetIntermediateALL:PRD112}. The aim of the present work is to close this gap for di-jet production, building on the extension of the \POWHEGBOX{} framework presented in \cite{Borsa:2024rmh}, which enables the simulation of processes with longitudinally polarized initial-state particles.

To this end, we present a Monte Carlo event generator for di-jet production in longitudinally polarized proton-proton collisions at NLO accuracy, implemented within the \POWHEGBOXVV{}, which allows consistent matching to parton showers. This implementation extends existing fixed-order calculations and enables realistic event simulations for polarized di-jet observables at RHIC. The paper is organized as follows:  in Sec.~\ref{sec:implementation} we describe the details of the implementation. Section~\ref{sec:pheno} is devoted to a detailed validation of the code and to phenomenological studies of recent dijet measurements by the STAR collaboration. Our conclusions are presented in Sec.~\ref{sec:conclusions}.
\section{Implementation}\label{sec:implementation}
\subsection{Jet observables in proton-proton collisions}

In this work, we study jet production in longitudinally polarized proton–proton collisions. In particular, we focus on dijet production,
\beq
\label{eq:pp_dijets_process}
p(p_1, \Lambda_1)+p(p_2, \Lambda_2)\rightarrow \mathrm{jet}(p^{\mathrm{jet\,1}})+\mathrm{jet}(p^{\mathrm{jet\,2}})+X,
\eeq
where $p_1,p_2$ and $\Lambda_1,\Lambda_2$ correspond, respectively, to the momenta and helicities of the incoming protons, while $p^{\mathrm{jet\,1}},p^{\mathrm{jet\,2}}$ are used to indicate the momenta of the reconstructed jets.

The differential cross section for a generic jet observable in polarized hadronic collisions can be factorized as \cite{Collins:1989gx}
\beq
\label{eq:xsec_factorization}
\begin{split}
d\sigma^{(\Lambda_1,\Lambda_2)}(p_1,p_2;\mathcal{S})=\sum_{a_1,a_2}\sum_{\lambda_1,\lambda_2}&\int dx_1 dx_2 \,f_{a_1\lambda_1}^{\Lambda_1}(x_1, \mu_F^2)\,f_{a_2\lambda_2}^{\Lambda_2}(x_2, \mu_F^2)\\&\times d\hat{\sigma}^{(\lambda_1,\lambda_2 )}_{a_1 a_2}(x_1 p_1, x_2 p_2; \mu_F^2; \mathcal{S}),
\end{split}
\eeq
with $d\hat{\sigma}_{a_1 a_2}$ denoting the partonic cross section, which now depends on the helicities and momenta of the partons involved in the hard-scattering process, determined by the momentum-fractions $x_1$ and $x_2$. In this case $f^{\Lambda_i}_{a_1\lambda_i}$ represents the helicity-dependent parton distribution function for a parton $a_i$ with helicity $\lambda_i$ inside a proton with helicity $\Lambda_i$, and depends both on the momentum fraction $x_i$ and the factorization scale $\mu_F^2$. The quantity $\mathcal{S}$ corresponds to the measurement function, which determines the jet momenta in terms of that of the outgoing partons.

 When considering unpolarized observables, the average over the helicities of the incoming hadrons is performed, in which case the cross section factorizes in the usual way in terms of a partonic cross section and standard unpolarized PDFs.

\beq
\label{eq:xsec_factorization_unpol}
\begin{split}
d\sigma(p_1,p_2;\mathcal{S})=&\frac{1}{4}\left[d\sigma^{(+,+)}+d\sigma^{(-,-)}+d\sigma^{(+,-)}+d\sigma^{(-,+)}\right]\\=&\sum_{a_1,a_2}\int dx_1 dx_2 \,f_{a_1}(x_1, \mu_F^2)\,f_{a_2}(x_2, \mu_F^2)\, d\hat{\sigma}_{a_1 a_2}(x_1 p_1, x_2 p_2; \mu_F^2; \mathcal{S}),
\end{split}
\eeq
where the unpolarized partonic cross section $d\hat{\sigma}_{a_1 a_2}$ is given by the analogue of the first line of Eq.~\eqref{eq:xsec_factorization_unpol}, 
\beq\label{eq:helicity_comb_unpol}
d\hat{\sigma}=\frac{1}{4}\left[ d\hat{\sigma}^{(+,+)}+ d\hat{\sigma}^{(-,-)}+ d\hat{\sigma}^{(+,-)}+ d\hat{\sigma}^{(-,+)}\right],
\eeq
and the unpolarized PDFs can be obtained from their helicity-dependent counterparts as 
\beq\label{eq:unpol_comb_pdfs}
f_{a_i}(x_i,\mu_F^2)=f_{a_i +}^{+}+f_{a_i -}^{+}=f_{a_i -}^{-}+f_{a_i +}^{-}.
\eeq

As for the case of polarized observables, instead of directly calculating the cross section from Eq.~\eqref{eq:xsec_factorization}, in practice it is often convenient to consider the longitudinally polarized cross section, defined in terms of the combination of helicity configurations
\beq\label{eq:helicity_comb}
d\Delta\sigma=\frac{1}{4}\left[d\sigma^{(+;+)}+d\sigma^{(-;-)}-d\sigma^{(+;-)}-d\sigma^{(-;+)}\right].
\eeq
Using Eq.~(\ref{eq:xsec_factorization}) it is possible to recast the longitudinally polarized cross section as
\beq
\label{eq:pol_fact}
d\Delta\sigma(p_1;p_2;\mathcal{S})=\sum_{a_1,a_2}\int dx_1 dx_2 \Delta f_{a_1}(x_1, \mu_F^2)\Delta f_{a_2}(x_2, \mu_F^2)\times d\Delta\hat{\sigma}_{a_1 a_2}(x_1 p_1; x_2 p_2; \mu_F^2;\mathcal{S}),
\eeq
where now $\Delta f_{a_1}(x_1)$ indicates the polarized PDF for parton $a_i$, corresponding to the difference between the helicity-dependent distributions for a parton with its helicity aligned and anti-aligned with that of the proton, that is,
\beq\label{eq:helicity_comb_pdfs}
\Delta f_{a_i}(x_i, \mu_F^2)=f_{a_i +}^{+}-f_{a_i -}^{+}=f_{a_i -}^{-}-f_{a_i +}^{-}.
\eeq
These polarized PDFs can be determined in global analyses of longitudinally polarized data in complete analogy to the case of unpolarized parton distribution functions~\cite{Ethier:2020way}. On the other hand,  $d\Delta\hat{\sigma}_{a_1 a_2}$ is the partonic analogue to Eq.~\eqref{eq:helicity_comb}, which can be calculated perturbatively in QCD.  In the particular case of di-jet production in polarized hadronic collision, the current state of the art corresponds to the NLO-QCD calculation of~\cite{deFlorian:1998qp}, which is based on the $2\rightarrow2$ and $2\rightarrow3$ helicity amplitudes obtained in \cite{Gastmans:1990xh} and \cite{Kunszt:1993sd}. In that work, a numerical implementation was obtained through a generalization of the FKS subtraction scheme \cite{Frixione:1995ms} that accounts for the initial-state polarization.


\subsection{Implementation in the \POWHEGBOX{}}

As stated in the introduction, the aim of our present work is to extend previous fixed-order calculations and develop a NLO parton-shower simulator for dijet production, which we implement in the \POWHEGBOXVV{}~\cite{Alioli:2010xd}. While the \POWHEG{} scheme~\cite{Nason:2004rx,Frixione:2007vw} was originally designed for unpolarized proton-proton collisions, the necessary modifications to extend the framework and simulate events in longitudinally polarized processes were discussed in detail in our previous work~\cite{Borsa:2024rmh}. In the following, we highlight some of the most salient points of that extension.

We start by considering the fixed-order NLO cross section. At NLO, and using a notation similar to that of the original \POWHEG{} paper~\cite{Frixione:2007vw}, the differential polarized cross section for a generic process involving $n$ external particles at leading order can be written as
\beq
\label{eq:pol_xsec}
d\Delta\sigma=\,d\psn{} \, \bigg[\Delta B(\psn)+ \Delta V(\psn)+ \int\,d\psrad\, \Delta R(\psn, \psrad) + \int dz \,\Delta G(\psn,z) \bigg],
\eeq
where $d\psn$ indicates the phase space of the underlying Born process, including the momentum fractions of the incoming partons. It is further assumed that the $(n+1)-$particle phase space can be factorized in terms of an underlying Born phase space and a radiation phase space~$d\mathbf{\Phi}_{\mathrm{rad}}$, according to
\beq
\label{eq:ps_factorization}
d\mathbf{\Phi}_{n+1}=d\mathbf{\Phi}_n\,d\mathbf{\Phi}_{\mathrm{rad}}\,.
\eeq
As for the remaining contributions in Eq.~(\ref{eq:pol_xsec})$, \Delta B$ corresponds to the  polarized Born contribution to the cross section, while $ \Delta V$ and $ \Delta R$ are the polarized virtual and real-emission corrections. The terms $\Delta G$ in Eq.~\eqref{eq:pol_xsec} denote the collinear factorization terms
which are introduced in hadronic cross sections to cancel some of the divergences associated with initial-state collinear emission. In these cases, $z$ denotes the fraction of momentum of the incoming parton after radiation. It should be noted that, in the notation employed in Eq.~\eqref{eq:pol_xsec}, all the previous contributions include the polarized PDFs and, therefore, the equation already includes the convolution between the partonic cross section and the polarized parton distribution functions. Furthermore, the sum over all possible partonic channels is assumed.

Following \cite{Borsa:2024rmh}, it is possible to extend the definition of the \POWHEG{} cross section to the polarized case as
\begin{equation}\label{eq:POWHEG_xsec_pol}
d\Delta\sigma_{\mathrm{PWHG}}=\Delta\overline{B}(\mathbf{\Phi}_n)\,d\mathbf{\Phi}_n\,\left\{\Delta^{\mathrm{pol}}(\mathbf{\Phi}_n, p_T^{\mathrm{min}})+\Delta^{\mathrm{pol}}(\mathbf{\Phi}_n, k_T(\mathbf{\Phi}_{n+1}))\frac{\Delta R(\mathbf{\Phi}_{n+1})}{\Delta B(\mathbf{\Phi}_{n})}\right\}, 
\end{equation}
where we have introduced the polarized \POWHEG{} Sudakov factor $\Delta^\pol$, which is defined in analogy to the unpolarized case \cite{Frixione:2007vw}, with the corresponding replacements of the real and Born contributions by  their polarized analogues, $\Delta R(\mathbf{\Phi}_{n+1})$ and $\Delta B(\mathbf{\Phi}_{n})$, i.e.\ 
\begin{equation}\label{eq:pol_POWHEG_sudakov}
    \Delta^\pol(\mathbf{\Phi}_n,p_T)= \exp\left\{ -\int\frac{\left[d\mathbf{\Phi}_{\mathrm{rad}} \,\Delta R(\mathbf{\Phi}_{n+1})\,\theta(k_T(\mathbf{\Phi}_{n+1})-p_T)\,\right]}{\Delta B(\mathbf{\Phi}_n)}\right\}.
\end{equation}
The function $k_T(\mathbf{\Phi}_{n+1})$ is used to measure the hardness of the emitted radiation and should be equal, near the infrared limits, to the transverse momentum of the emitted parton.  The parameter $p_T^{\mathrm{min}}$  corresponds to an infrared cut-off, below which  real radiation is considered to be unresolved, which is typically chosen to be of order 1 GeV.
The \POWHEG{} function $\overline{B}(\mathbf{\Phi}_n)$, on the other hand, is defined as
\beq\label{eq:POWHEG_btilde_pol}
\begin{split}
\Delta \overline{B}(\mathbf{\Phi}_n)=& \Delta B(\mathbf{\Phi}_n)+ \Delta V(\mathbf{\Phi}_n)+\left[\int d\mathbf{\Phi}_{\mathrm{rad}} \Delta C(\mathbf{\Phi}_{n+1})+  \int dz\, \Delta G(\mathbf{\Phi}_{n},z)\right]\\ 
&+\left[\int d\mathbf{\Phi}_{\mathrm{rad}} \left[\Delta R(\mathbf{\Phi}_{n+1}) - \Delta C(\mathbf{\Phi}_{n+1})\right] \right].
\end{split}
\eeq
The quantities $\Delta B$, $\Delta V$, $\Delta R$ and $\Delta G$ are the same as in Eq.~\eqref{eq:pol_xsec}, and $\Delta C$ is used to indicate the infrared counter-terms introduced to cancel the divergences stemming from the real-emission corrections at the integrand level. As in the unpolarized case, all the divergent contributions present in Eq.~\eqref{eq:POWHEG_btilde_pol} cancel, resulting in a finite value of $\Delta \overline{B}$.

Based on Eq.~\eqref{eq:POWHEG_xsec_pol}, the generalization of the unpolarized dijet implementation of Ref.~\cite{Alioli:2010xa}  to account for the polarization of initial-state particles requires a consistent replacement of the quantities, $\bbar$, $R$,  $B$, and the Sudakov factor $\Delta$ with their polarized counterparts. 
In the case of $\Delta B$ and $\Delta R$, the relevant contributions can be directly obtained from the helicity amplitudes for $2\rightarrow2$ and $2\rightarrow3$ processes available in the literature \cite{Gastmans:1990xh}, which then only need to be combined with appropriate factors according to Eq.~\eqref{eq:helicity_comb}.
Similarly, the necessary helicity amplitudes for the one-loop contributions can be obtained from \cite{Kunszt:1993sd}. In the case of the virtual helicity contributions, we have used the {\tt Fortran} routines from the Monte-Carlo code developed in Ref.~\cite{deFlorian:1998qp}. 
In principle, it is also possible to use the \MADGRAPH{} amplitude generator~\cite{Alwall:2011uj} and the {\tt MadLoop} module \cite{Hirschi:2011pa} to automatically generate all the Born, real and virtual contributions for dijet production. Since in \MADGRAPH{}  those contributions are constructed from  helicity amplitudes generated via the {\tt ALOHA} \cite{deAquino:2011ub} libraries, they can also be simply combined to produce the contributions in Eq.~\eqref{eq:POWHEG_xsec_pol}. As it will be explained in section \ref{sec:pheno}, this option was also explored as an additional check on the correct implementation of the process.

The remaining modifications to implement Eq.~\eqref{eq:POWHEG_xsec_pol} within the standard \POWHEGBOX{} are process-independent, and related to the generation of the infra-red counterterms and the finite remnants of the collinear factorization. As explained in detail in \cite{Borsa:2024rmh}, the former are addressed by implementing the polarized extension of the FKS subtraction scheme developed in  \cite{deFlorian:1998qp, deFlorian:1999ge}. As for the latter, it is enough to replace the unpolarized Altarelli-Parisi splitting functions by their polarized analogues \cite{Vogelsang:1996im}.

It should be noted that, since the polarized cross section might involve negative weights even at leading order, we use the flag {\tt withnegweights} to handle potentially negative contributions. In that case, the \POWHEGBOX{} generates both positive- and negative-weight events with probabilities $n_+$ and $n_-$ given by
\beq
\label{eq:event_fraction}
n_+=\frac{\sigma_{(+)}}{\sigma_{(+)}+|\sigma_{(-)}|}\,,\qquad 
n_-=\frac{|\sigma_{(-)}|}{\sigma_{(+)}+|\sigma_{(-)}|},
\eeq
In the previous equation,  $\sigma_{(+)}$ and $\sigma_{(-)}$ denote the positive and negative contributions to the \textit{total} cross section, respectively. The positive- and negative-weight events are then assigned fixed weights of  $\pm\left(\sigma_{(+)}+|\sigma_{(-)}|\right)$, so that the mean value of the cross section is given by
\beq
\sigma = +\left(\sigma_{(+)}+|\sigma_{(-)}|\right) n_+ - \left(\sigma_{(+)}+|\sigma_{(-)}|\right) n_- = \sigma_{(+)} - |\sigma_{(-)}| = \sigma_{\mathrm{NLO}},
\eeq

and therefore the NLO accuracy for the inclusive cross section is retained.

\section{Phenomenological results}
\label{sec:pheno}
In the remainder of this section, we study the phenomenology of di-jet production in longitudinally polarized proton-proton collisions at RHIC center-of-mass-system (c.m.s.) energies of $\sqrt{s}=510~\text{GeV}$ and $\sqrt{s}=200~\text{GeV}$, given by the reaction
\beq
p+p\rightarrow \mathrm{jet}_1(p^{\mathrm{jet\,1}})+\mathrm{jet}_2(p^{\mathrm{jet\,2}})+X,
\eeq
with $\mathrm{jet}_1,\,\mathrm{jet}_2$ denoting the two hardest jets produced in the event, and $p^{\mathrm{jet\,1}},\,p^{\mathrm{jet\,2}}$ their corresponding momenta. Unless stated otherwise, the momenta of the jets are reconstructed using the anti-$k_T$ algorithm \cite{Cacciari:2008gp} and $E-$recombination scheme, with $R=1$. In the context of the \PBOX{} this is done using the {\tt FastJet} library~\cite{Cacciari:2011ma}. Furthermore, we set the values of the factorization and renormalization scales $\mu_F$ and $\mu_R$ according to 
\beq
\label{eq:scales}
\mu_F=\xif\mu_0,\quad \mu_R=\xir\mu_0,\quad\text{where}\quad\mu_0=p_T,
\eeq 
and for each simulated event $p_T$ corresponds to the transverse momentum of the underlying Born configuration, that is, to the transverse momentum of any of the two partons produced back-to-back in the underlying $2\rightarrow2$ event. In order to assess scale uncertainties, the factors $\xif$ and $\xir$ will be varied from 0.5 to 2 under the constraint $1/2\leq\xif/\xir\leq2$. 

In all cases, we use an active number of flavors $N_f=4$, the NLO set of polarized PDFs from {\tt BDSSV24}\cite{Borsa:2024mss} and its corresponding value of $\alpha_S$ as implemented in the {\tt LHAPDF6} library~\cite{Buckley:2014ana}. We have additionally checked that the presented results are consistent within their intrinsic uncertainties  when considering the set of polarized PDFs from \cite{Cruz-Martinez:2025ahf}.
\subsection{Validation} 
Before analyzing the phenomenological impact of our event generator in polarized proton-proton collisions, we dedicate this section to a validation of its implementation in the \POWHEGBOX{}. For this purpose, we first perform a detailed comparison of our fixed-order results to those previously available in the literature. Additionally, we analyze the hardest-emission events generated by {\POWHEG}, and finally study the impact of matching to a parton shower.

As a starting point, we consider the fixed-order NLO cross-section for di-jet production at c.m.s energy of $\sqrt{s}=510~\text{GeV}$. While events generated by the \POWHEGBOX{} are not necessarily distributed according to the NLO cross-section, the latter plays an important role within \POWHEG{}, as it determines the probability distribution for the underlying Born events, from which the hardest-radiation events are subsequently generated, through the $\overline{B}$ function or its polarized analogue $\Delta\overline{B}$.

To validate our implementation, we study representative jet distributions and compare the fixed-order results obtained with our default \PBOX{} implementation to those from the NLO code of \cite{deFlorian:1998qp}, which will be labelled as {\tt dFFSV}. As an additional check, we also present results obtained with an alternative \PBOX{} implementation  using matrix elements generated by \MADGRAPH{}.  In this case, \MADGRAPH{} is used to produce the spin- and color-correlated Born amplitudes, the real-emission contributions, and the finite part of the virtual corrections. The relevant matrix elements can be directly generated in the necessary \POWHEG{} format using  the interface to \MGAMCNLO{} \cite{Alwall:2014hca} developed in \cite{Nason:2020lxx}, which we have further modified  to account for the polarization of initial-state particles \cite{Betz2025}.
Since matrix elements within {\MADGRAPH} are calculated from individual helicity amplitudes  (originally obtained with the {\tt HELAS} package \cite{Murayama:1992gi} and more recently by the {\tt ALOHA} package \cite{deAquino:2011ub}), this simply amounts to properly adjusting the signs for each of the different helicity configurations according to Eq.~\eqref{eq:helicity_comb}. 

For these comparisons, we employ fixed scales setting $\mu_F=\mu_R=50~\text{GeV}$.
Moreover, a generation cut of $p_T > 3~\text{GeV}$ is imposed  on the transverse momenta of partons in the underlying Born events and, within the \PBOX{} a Born suppression factor of the form 
\beq
S(p_T)=\left(\frac{p_T^2}{p_T^2+p_{T,\mathrm{supp}}^2}\right)^3\,,
\eeq
with $p_{T,\mathrm{supp}}=20~\text{GeV}$ is applied.
We have checked that reducing either the generation cut or the parameter $p_{T,\mathrm{supp}}$ does not modify the presented results. 

%
\begin{figure}[tp]
    \centering
    \includegraphics[width=0.83\textwidth]{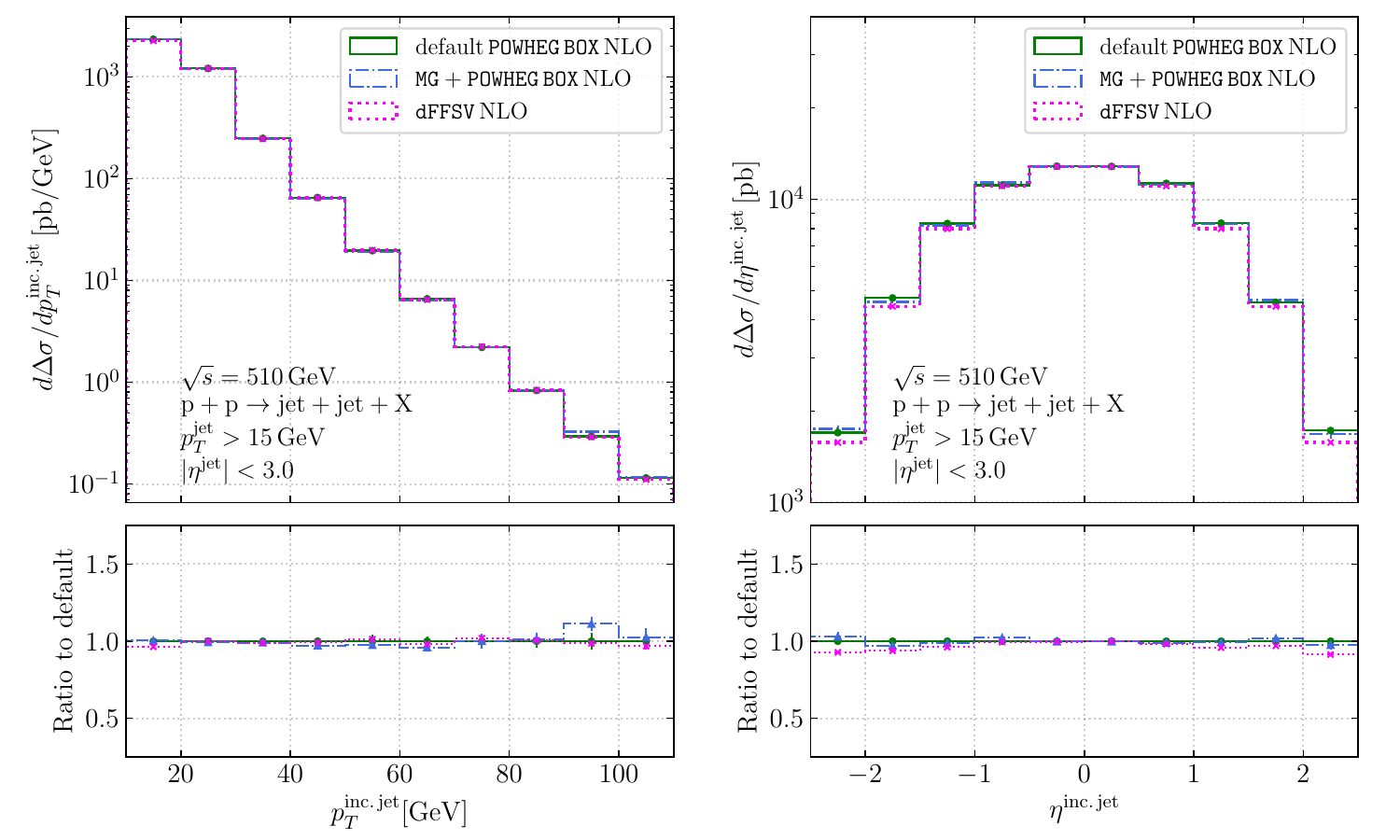}%

    \medskip
    \centering
    \includegraphics[width=0.83\textwidth]{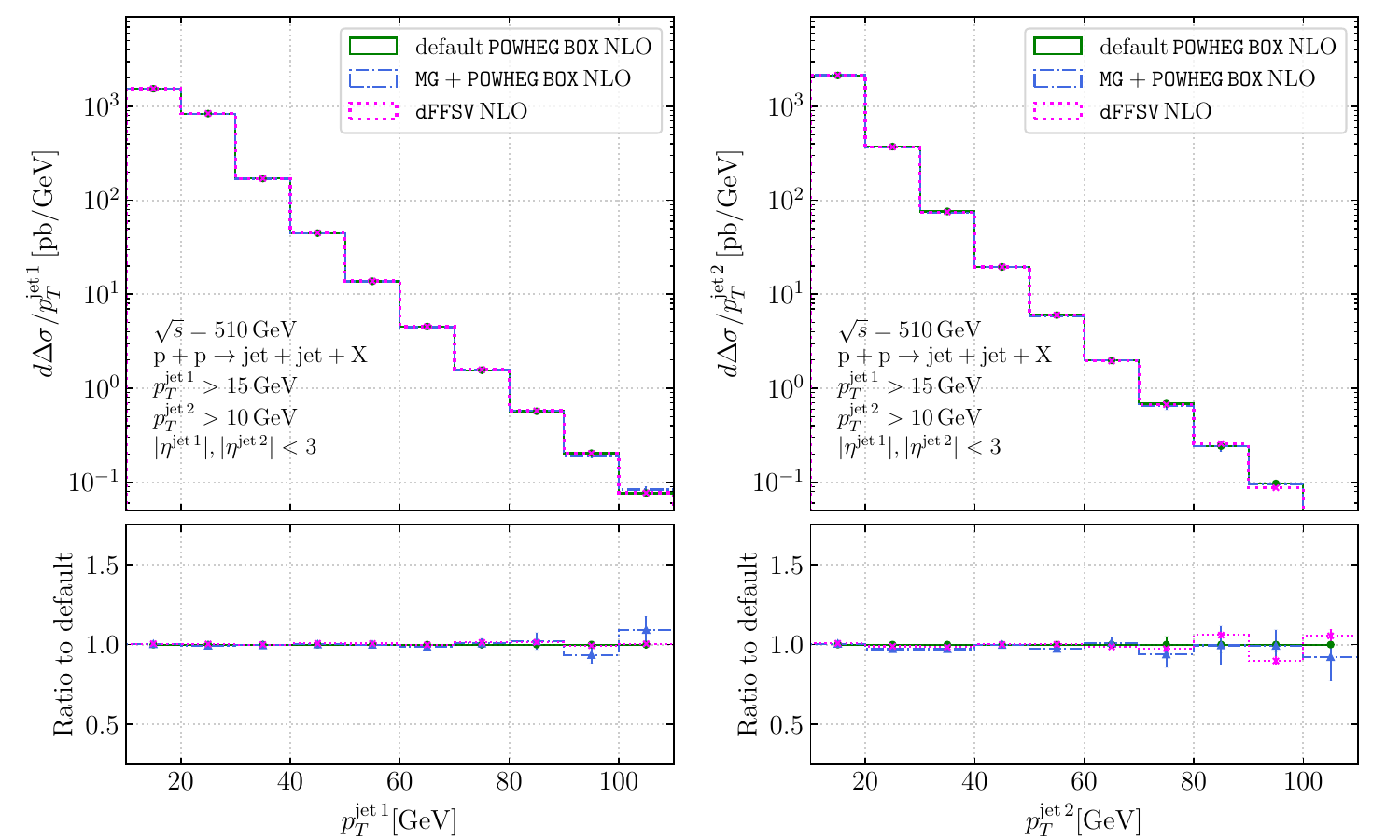}%
    \caption{
    \label{fig:NLOcomp}
    NLO differential cross section  as a function of the inclusive-jet transverse momentum $p_T^{\text{inc.\,jet}}$ (top left), pseudo-rapidity  $\eta^{\text{inc.\,jet}}$ (top right), and the transverse-momenta of the two hardest jets  $\ptjone$, $\ptjtwo$ (bottom left and right, respectively) for di-jet production at $\sqrt{s}=510~\text{GeV}$, for the cuts indicated in the respective panels, as obtained with our default \POWHEGBOX{}  (solid green) implementation, {\tt dFFSV} code (dotted pink) and a  \MADGRAPH{}-based \POWHEGBOX{} implementation (dot-dashed blue). In each case, the lower panels depict the ratio to the result from the standard \PBOX{} implementation. Error bars indicate the numerical uncertainty.
    }
\end{figure}
%

Figure~\ref{fig:NLOcomp} presents a comparison of the results of the three programs for the inclusive-jet transverse momentum $p_T^{\text{inc.\,jet}}$ and pseudo-rapidity $\eta^{\text{inc.\,jet}}$ distributions, as well as for the transverse momentum  distributions of the two hardest jets, $\ptjone$ and $\ptjtwo$.  In the case of inclusive-jet distributions, we bin simultaneously the two leading jets, provided they satisfy  
\beq
\label{eq:inc-cuts}
p_T^{\text{jet}}>15~\text{GeV}, \quad  
|\eta^{\text{jet}}|<3 
\eeq
in events with at least two jets. The distributions in the transverse-momenta of the two hardest jets are obtained applying the selection cuts  
\beq
\label{eq:star-cuts}
\ptjone>15~\GeV, \qquad \ptjtwo>10, \qquad |\etajone|,|\etajtwo|<3.
\eeq 
 In all cases, as well as for other distributions not shown here,  we find the results from the different codes to be fully consistent within their numerical uncertainties.

%
\begin{figure}[tp]
    \centering
    \includegraphics[width=0.9\textwidth]{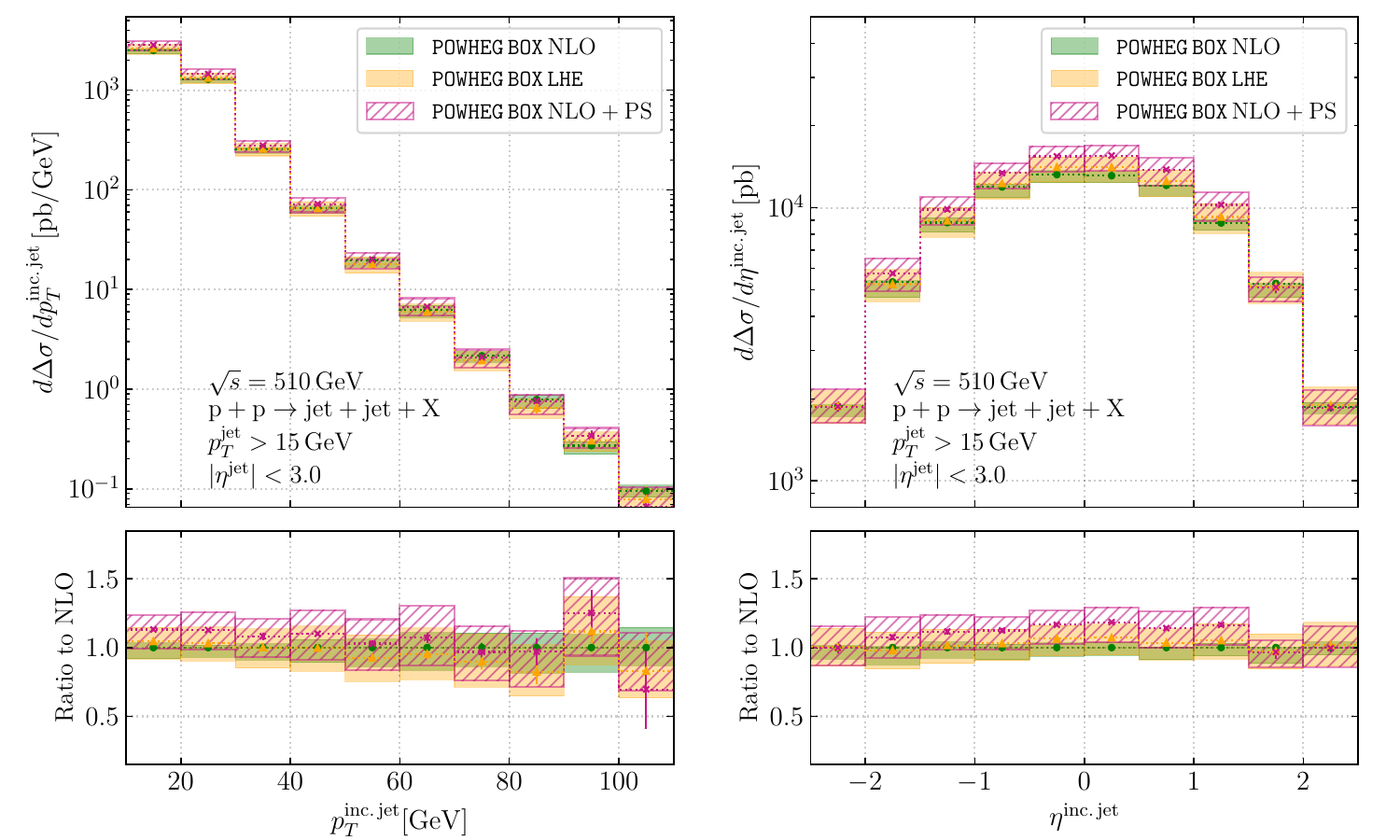}%
    \caption{
    \label{fig:LHEcomp_inc}
    Differential cross section for di-jet production at $\sqrt{s}=510~\text{GeV}$ as a function of the inclusive-jet transverse momentum $p_T^{\text{inc.\,jet}}$ (left) and pseudo-rapidity  $\eta^{\text{inc.\,jet}}$ (right) at NLO (green), {\tt LHE} (yellow) and NLO+PS (pink) level, as obtained with our \POWHEGBOX{} implementation. The lower panels display the ratio to the fixed-order NLO result. Error-bars indicate the numerical uncertainty from the Monte Carlo integration, while the bands indicate  the theoretical uncertainty from the 7-point scale variation.
    }
\end{figure}
%

Having validated the fixed-order NLO implementation, we now turn to the analysis of the hardest-emission radiation generated in \POWHEG{}. Within this scheme, the first branching is generated from an underlying Born event according to Eq.~\eqref{eq:POWHEG_xsec_pol} which, in addition to the pure NLO contributions, involves the all-order resummation of enhanced contributions in the soft and collinear limits through the presence of the Sudakov form factor. As a result, the distributions obtained at this stage can deviate from the fixed-order NLO prediction in regions where Sudakov effects are significant. 
To assess these effects, we compare the NLO distributions from Fig.~\ref{fig:NLOcomp} with those corresponding to the hardest-emission events generated by \POWHEG{}, which can be obtained from the Les Houches Events (LHE) files produced within the \POWHEGBOX{} prior to the application of the parton shower. In the following, we refer to these results as {\tt LHE} distributions. Figure~\ref{fig:LHEcomp_inc} presents  the comparison between the results from the \PBOX{} at the NLO and the {\tt LHE} level for the inclusive-jet distributions previously considered in Fig.~\ref{fig:NLOcomp}. In this case, the bands represent the theoretical uncertainty estimated through the standard 7-point scale variation, as indicated in Eq.~\eqref{eq:scales}.

It should be noted that, for distributions that  are inclusive in the additional radiation, and therefore insensitive to Sudakov effects, a good level of agreement between the hardest-emission distributions from {\tt POWHEG} and the fixed-order results is expected \cite{Frixione:2007vw},  with deviations smaller than the expected higher-order corrections. This is precisely the case for the distributions in the inclusive-jet transverse momentum and pseudo-rapidity, which involve selection cuts only for the two hardest jets in an event and impose no constraints  on possible additional jets. It is therefore not surprising that these distributions show only small deviations between the NLO and {\tt LHE} results, with differences that are typically lower than $5\%$ and well within the scale uncertainty. 

%
\begin{figure}[tp]
    \centering
    \includegraphics[width=0.9\textwidth]{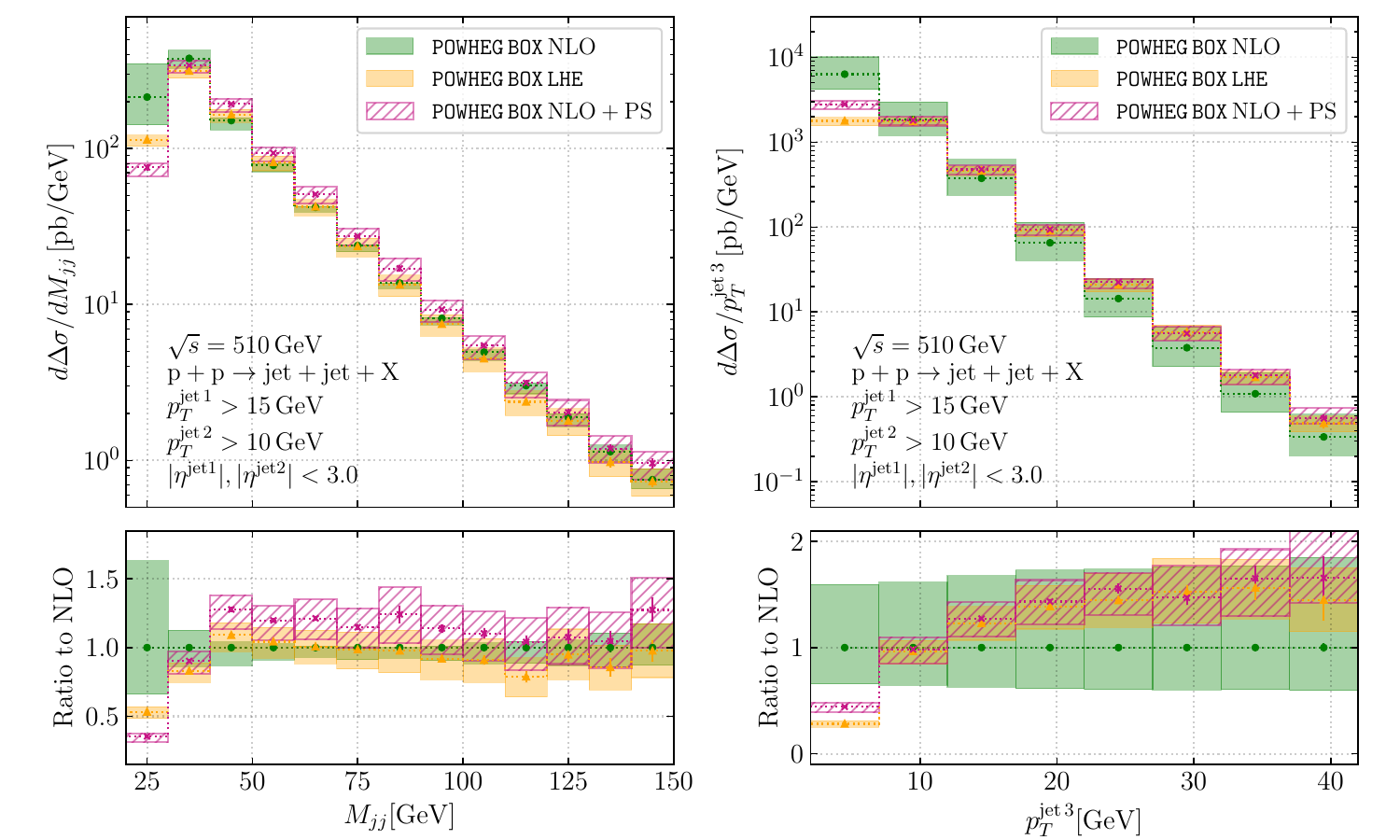}%
    \caption{
    \label{fig:LHEcomp_Mjj}
    Similar to Fig.~\ref{fig:LHEcomp_inc} for the invariant mass of the di-jet system $\mjj$ (left) and the transverse momentum of the third hardest jet $\ptjthree$ (right). The cuts of  Eq.~\eqref{eq:star-cuts} were applied. 
    }
\end{figure}
%

More exclusive jet distributions can, on the other hand, exhibit larger differences between the NLO and \POWHEG{} hardest-emission distributions. This can be observed, for example, in Fig.~\ref{fig:LHEcomp_Mjj}, which presents a comparison of the \POWHEG{} differential cross sections at the NLO and {\tt LHE} level as a function of the invariant mass of the di-jet system, $\mjj$, and the transverse momentum  of the third hardest jet, $\ptjthree$, in a fashion similar to Fig.~\ref{fig:LHEcomp_inc} and using the selection cuts of Eq.~\eqref{eq:star-cuts}. Being inclusive in the emission of additional partons, the distribution in the invariant-mass of the di-jet system also shows good agreement with the NLO result in the intermediate-to-high-$\mjj$ region, where Sudakov effects are expected to be small, while a strong suppression in the first couple of bins is observed. This behavior can be understood by noting that, due to the selected kinematical cuts, the region $\mjj<30~\text{GeV}$ is kinematically forbidden at LO and only gets populated starting at NLO. For those first few bins, the fixed-order result is then effectively only LO-accurate, and higher-order QCD corrections are expected, as evidenced by the increase in the scale dependence in the low-$\mjj$ region. These bins should therefore be interpreted with caution and are not expected to provide precision phenomenological information at the NLO level.
As for the distribution in $\ptjthree$, it should be noted that, in the absence of parton showering, events with three jets are necessarily generated from the real emission contributions. In that case, the difference observed in the first bin is expected: while the NLO cross section diverges as $\ptjthree\rightarrow 0$, corresponding to the limit of soft or collinear emission, the {\tt LHE} distributions are subject to a Sudakov damping, leading to the sharp suppression observed in the first bin. While sizable, the differences in the remaining range of $\ptjthree$ are still within the scale-uncertainty bands, which are large due to the fact that the distribution is only LO accurate. 

%
\begin{figure}[tp]
    \centering
    \includegraphics[width=0.9\textwidth]{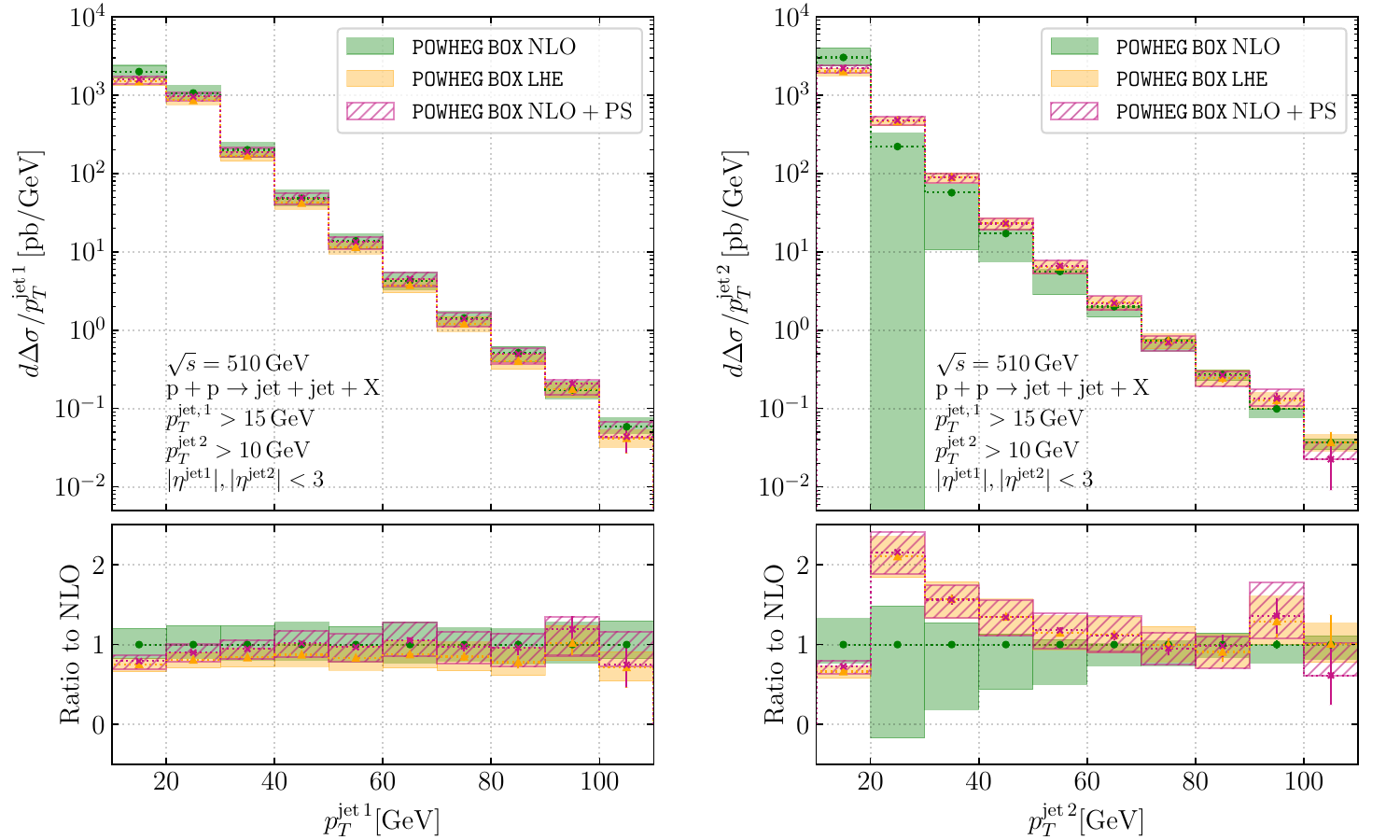}%
    \caption{
    \label{fig:LHEcomp_pt}
    Similar to Fig.~\ref{fig:LHEcomp_Mjj} for the transverse momenta of the two hardest jets.
    }
\end{figure}
%
%
For other exclusive distributions, the differences between NLO and {\LHE} distributions can be even more pronounced. As an example, in Fig.~\ref{fig:LHEcomp_pt} we consider the hardest-emission cross sections as distributions in the transverse-momenta of the two hardest jets. 
Compared with the cross section as a function of $\ptjthree$, the distribution in the transverse momentum of the hardest-jet is less affected by Sudakov effects and exhibits a milder difference between the NLO and hardest-emission cross sections, with the latter presenting a suppression of $15-20 \%$ in the low-$\ptjone$ region, compatible with the uncertainty from missing higher-order corrections.

In contrast, the behavior of the distribution in the transverse momentum of the second-hardest jet is more involved. In this case, the {\tt LHE} results present an excess of $50\%-100\%$ in the region $20~\mathrm{GeV}\lesssim \ptjtwo\lesssim 60~\mathrm{GeV}$ compared to the fixed-order result. We have observed a similar feature in unpolarized di-jet production at comparable energies.
This stark difference between the hardest-emission and fixed-order predictions can be traced back to the fact that the latter becomes unreliable in the near back-to-back configuration, 
$\frac{\ptjone-\ptjtwo}{\ptjone+\ptjtwo}\ll1$, 
where Sudakov logarithms can lead to large negative contributions at fixed-order \cite{Chen:2016cof, Mueller:2015ael}. This is precisely the case for the bins in the range $20~\mathrm{GeV}\lesssim \ptjtwo\lesssim 60~\mathrm{GeV}$. The negative corrections in this region lead to the increase of the scale-uncertainty bands, and can even result in negative values of the cross section for some of the scale choices, as is the case in the second bin. 
These logarithmically enhanced contributions are regulated in the hardest-emission generation of \POWHEG{} through the inclusion of a Sudakov suppression factor, leading to a smoother increase of the cross section.
%
%
\begin{figure}[tp]
    \centering
    \includegraphics[width=0.6\textwidth]{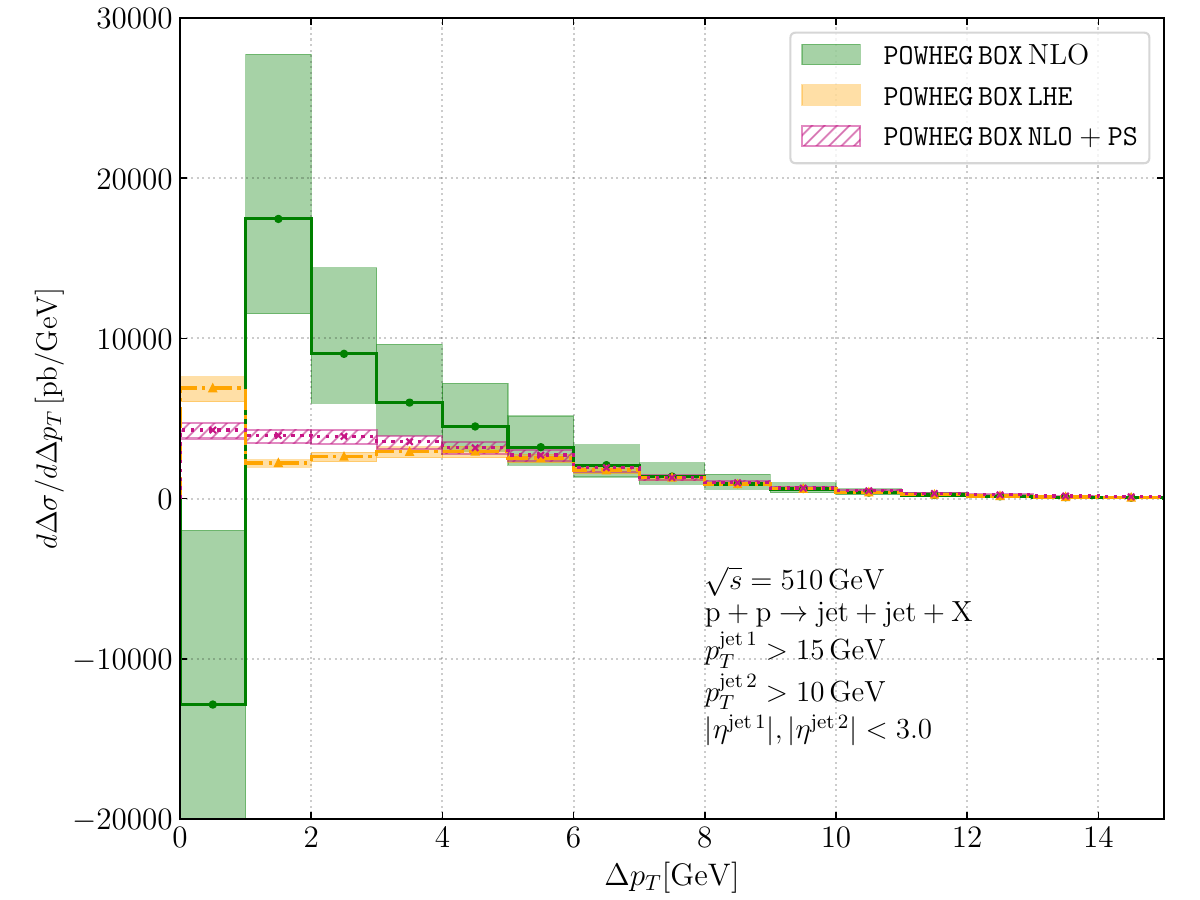}%
    \caption{
    \label{fig:LHEcomp_Delta_pt} Inclusive jet cross section as a function of $\Delta p_T=(\ptjone-\ptjtwo)$ for jets satisfying the same cuts as in Fig.~\ref{fig:LHEcomp_pt} at the NLO (green), {\tt LHE} (yellow), and NLO+PS (pink) levels.}
\end{figure}
%
%
To illustrate this point, Fig.~\ref{fig:LHEcomp_Delta_pt} presents the cross section as a function of the relative transverse momentum between the two hardest jets, $\Delta p_T=\ptjone-\ptjtwo$, using the same cuts as those employed in Fig.~\ref{fig:LHEcomp_pt}. As can be seen, the fixed-order cross section increases as lower values of $\Delta p_T$ are approached, with the exception of the first bin, corresponding to $\ptjone\approx \ptjtwo$, where it exhibits a sharp decrease and becomes negative. This pathological behavior originates from the interplay between real-emission contributions and subtraction counterterms in the back-to-back region: real radiation populates bins with $\Delta p_T>0$, while the corresponding negative-weighted counterterms retain Born kinematics and therefore contribute at $\Delta p_T=0$. The cancellation between these two contributions is highly sensitive to the rate of soft emission, which is not properly captured by the fixed-order cross section. The inclusion of the Sudakov suppression factor in the \POWHEG{} prediction, on the other hand, leads to a smoother behavior of the cross section at low $\Delta p_T$, effectively regulating the large negative fixed-order contributions and resulting in the observed excess in the {\tt LHE} distribution for $\ptjtwo$. It should also be noted that, by construction, the second hardest jet is more sensitive to soft radiation and to migrations induced by additional emissions, resulting in the relatively large effect compared to the distribution in $\ptjone$. A similar behavior for the $\Delta p_T$ distribution can also be observed in unpolarized di-jet production at higher energies \cite{Alioli:2010xa}.

Besides understanding the origin of the differences between the NLO and hardest-emission results, the analysis carried out in Fig.~\ref{fig:LHEcomp_Mjj} and Fig.~\ref{fig:LHEcomp_pt} highlights the potential shortcomings of fixed-order predictions in some kinematical regions when considering RHIC c.m.s.~energies. In that sense, it should be noted that these differences can be enhanced if low cuts for $\ptjone$ and $\ptjtwo$, as those employed in \cite{STAR:2025:DijetIntermediateALL:PRD112,STAR:2021mqa}, are considered.

Finally, after analyzing the hardest-emission generated by \POWHEG{}, we turn our attention to the effects of supplementing the NLO calculation with a parton shower, to which from now on we will refer to as NLO+PS. To this end, we analyze the cross section using exclusively the default shower phase of \PYTHIA{}, that is, de-activating hadronization and multi-parton-interaction effects. Specifically we use \PYTHIAS{} and the Perugia 2012 tune~\cite{Skands:2010ak}, which is the event generator employed in the STAR measurements we are ultimately interested in studying \cite{STAR:2021mqa, STAR:2025:DijetIntermediateALL:PRD112}. It is important to note that, since there are no publicly available parton shower codes for longitudinally polarized processes available, the shower stage of our predictions is necessarily based on helicity-averaged codes. As argued in \cite{Borsa:2024rmh}, even while using an unpolarized shower, most of the leading logarithmic contributions in the polarized process are still correctly captured.

The results from interfacing the fixed-order result with the parton shower are also depicted in Figs.~\ref{fig:LHEcomp_inc}, \ref{fig:LHEcomp_Mjj}, \ref{fig:LHEcomp_pt} and \ref{fig:LHEcomp_Delta_pt}  as dashed-pink histograms. As can be seen, there is in general a good level of agreement between the hardest-emission cross sections and their showered counterparts. In the case of Fig.~\ref{fig:LHEcomp_Mjj} and Fig.~\ref{fig:LHEcomp_pt}, compared to the hardest-emission cross section, the inclusion of the parton shower has a rather mild effect on the transverse momentum   distributions of the three hardest jets, with a slight increase in the distribution of the leading jet of around $10\%$ that brings the result closer to the fixed-order one. The {\tt LHE} distributions in $\ptjtwo$ and $\ptjthree$ are largely unaffected by the parton shower, with the most prominent effect being a reduction in the Sudakov suppression for the first bin of the $\ptjthree$ distribution. In the case of the more inclusive distributions, like those presented in Fig.~\ref{fig:LHEcomp_inc} and the invariant-mass distribution of Fig.~\ref{fig:LHEcomp_Mjj}, the effect of the interface with {\tt PYTHIA} is slightly more pronounced, with an enhancement of approximately $15\%$ for the inclusive-jet distributions in the low-$p_T$ and central-pseudo-rapidity regions. As for the invariant-mass distribution, we observe a slight shift towards larger values of $\mjj$. 

 These differences can ultimately be associated with the specific value of the jet-resolution parameter $R$ selected for the analysis. 
We have observed that, while we find good agreement between the NLO and {\tt LHE} results for inclusive distributions using our standard choice of $R=1$, this value of the resolution parameter also leads to the observed increase of the NLO+PS results. Lower values of $R$, on the other hand, can improve the agreement between the NLO and NLO+PS results, although producing more sizable differences with the {\tt LHE} distributions. This is related to the fact that, compared to the fixed-order simulation, the \POWHEG{} {\tt LHE} result has a reduced sensitivity to the choice of the jet resolution parameter $R$, as readily explained in the \PBOX{} implementation for unpolarized di-jet production of Ref.~\cite{Alioli:2010xa}.
%
%
\begin{figure}[tp]
    \centering
    \includegraphics[width=0.6\textwidth]{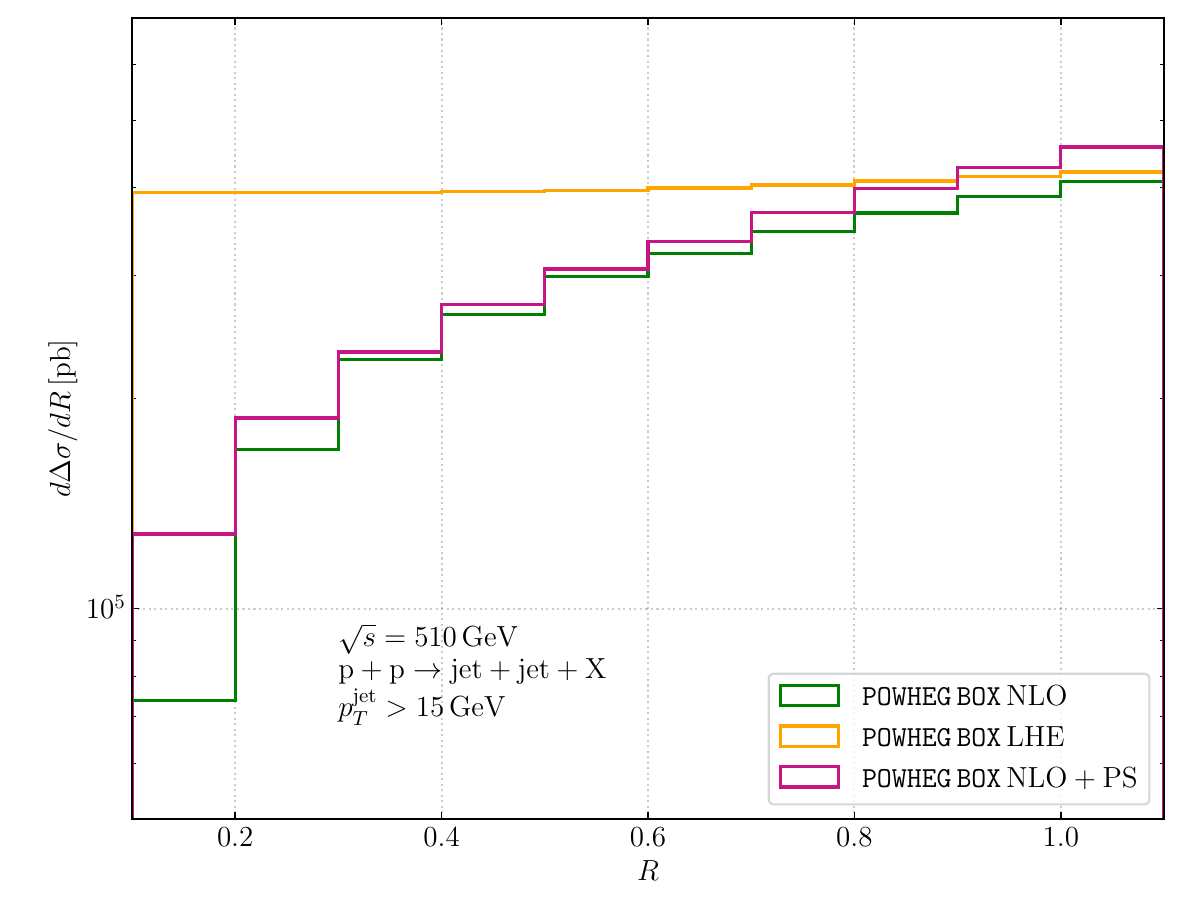}%
    \caption{
    \label{fig:LHEcomp_R}
    Inclusive jet cross section as a function of the jet resolution parameter $R$ for jets with transverse momenta larger than 15~GeV at the NLO (green), {\tt LHE} (yellow) and NLO+PS (pink) levels.
    }
\end{figure}
%
%
To highlight this point, and following the argument of~\cite{Alioli:2010xa}, Fig.~\ref{fig:LHEcomp_R} presents a comparison of the inclusive-jet cross section at the NLO, {\tt LHE} and NLO+PS levels as a function of the jet parameter $R$, requiring each jets' transverse momentum to be larger than $15~\mathrm{GeV}$. While the NLO results exhibit the expected divergent behavior as $R\rightarrow0$, the hardest-emission cross section from \POWHEG{} (i.e.\ the \LHE{} result), is largely unaffected by a reduction of $R$. The reason for this reduced sensitivity to the jet parameter is twofold: On the one hand, it is related to the way in which the first branching is produced within the \POWHEG{} scheme, with only one of the external particles being involved in the branching, and therefore reducing the dependence on $R$ that additional emission from the remaining external legs would produce. On the other hand, the presence of the Sudakov factor further suppresses  additional small-angle radiation thus reducing the dependence on $R$. The addition of the parton shower counteracts these effects, resulting in NLO+PS cross sections that are closer to the NLO ones. It should be noted that, for larger choices of the jet parameter $R$, the NLO and {\tt LHE} cross section start showing better agreement, while for $R\gtrsim0.8$, the parton shower actually results in an increase of the total cross section, leading to the behavior observed in Fig.~\ref{fig:LHEcomp_inc}.
%
%
\begin{figure}[tp]
    \centering
    \includegraphics[width=0.6\textwidth]{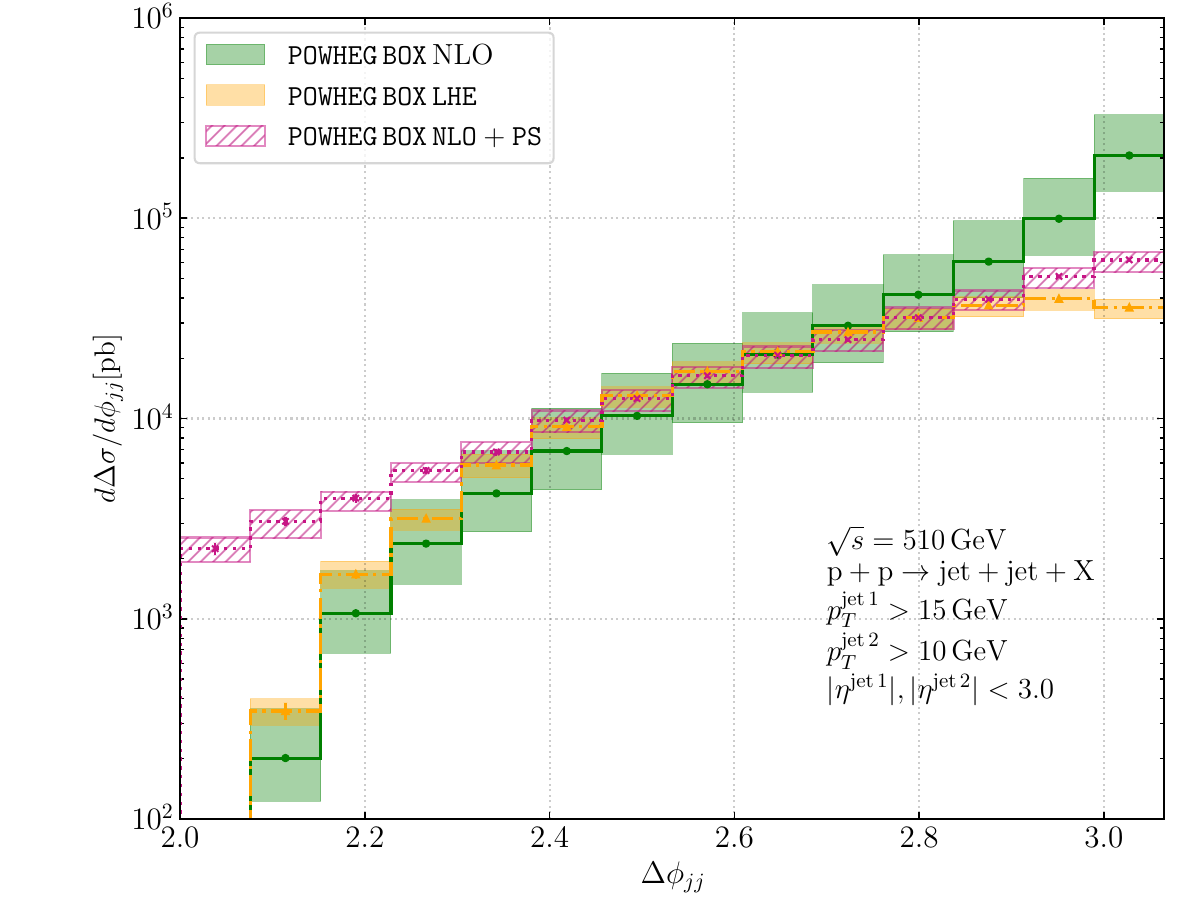}%
    \caption{
    \label{fig:delta_phi}
    Similar to Fig.~\ref{fig:LHEcomp_inc}, for the difference in azimuthal angle between the two leading jets,  $\phij$.
    }
\end{figure}
%

As a closing remark on the impact of parton-shower effects, it is worth emphasizing that, for observables sensitive to the emission of multiple partons, the showered predictions can differ significantly from those obtained at fixed order and at the hardest-emission level. As an example, Fig.~\ref{fig:delta_phi} shows a comparison of these three predictions for the azimuthal-angle separation of the two leading jets, $\phij$.

The behavior in the region $\phij \approx \pi$ is qualitatively similar to that previously observed for $\ptjthree$ and $\Delta p_T$ in Figs.~\ref{fig:LHEcomp_Mjj} and \ref{fig:LHEcomp_Delta_pt}. At the Born level, the two leading jets are produced strictly back-to-back, such that $\phij = \pi$. As a result, this region is highly sensitive to soft radiation, leading to large logarithmic corrections in the fixed-order prediction. These contributions are suppressed in both the {\tt LHE} and NLO+PS results by the presence of the Sudakov factor, yielding well-behaved distributions in the $\phij \to \pi$ limit.

In addition, sizable differences are observed at smaller values of $\phij$. While the NLO and {\tt LHE} predictions exhibit a sharp decrease as $\phij$ is reduced, the showered result shows a noticeably softer behavior. This can be understood by noting that configurations with two leading jets separated by a small azimuthal angle require recoil against additional emitted partons. Such configurations are generated only once multiple emissions are taken into account, as is the case in the parton-shower simulation. Since the NLO and {\tt LHE} predictions include at most a single additional emission, this region of phase space is correspondingly suppressed in those calculations.

\subsection{RHIC phenomenology}

After validating the results produced by our \POWHEGBOX{} implementation, we now turn to a phenomenological comparison of our predictions with experimental measurements performed by the STAR collaboration at RHIC. The STAR collaboration has reported several measurements of the longitudinal double-spin asymmetry $A_{LL}$, corresponding to the ratio of the polarized and unpolarized cross sections differential in an observable $\mc{O}$, i.e.\  
\beq
A_{LL} = \frac{d\Delta\sigma/d\mc{O}}{d\sigma/d\mc{O}}\,, 
\eeq
for inclusive-jet and di-jet-production in polarized proton-proton collisions at c.m.s.~energies of  $\sqrt{s}=510~\mathrm{GeV}$ and $\sqrt{s}=200~\mathrm{GeV}$, and different values of pseudo-rapidity~\cite{STAR:2006:JetALL:PRL97, STAR:2008:JetALL:PRL100, STAR:2012hth, STAR:2014wox, STAR:2016kpm, STAR:2018yxi, STAR:2019yqm, STAR:2021mfd, STAR:2021mqa, STAR:2025:DijetIntermediateALL:PRD112}.
In the following we consider their most recent measurements for di-jet-production~\cite{STAR:2021mqa,STAR:2025:DijetIntermediateALL:PRD112}.

In the case of the di-jet-measurements at $\sqrt{s}=510~\mathrm{GeV}$~\cite{STAR:2021mqa}, the double-spin asymmetry in the central region $|\etaj|<0.9$  is reported as a function of the invariant mass of the di-jet-system, for four different jet-pair topologies corresponding to the configurations given in Tab.~\ref{tab:star-tops}. 
\begin{table}[t]
\begin{center}
\begin{tabular}{ | c | c | c |}
\hline
 Topology A & $0.3<|\etajone|,|\etajtwo|<0.9;\quad\etajone\cdot\etajtwo>0$ & Forward-forward \\ 
 Topology B & $|\eta^{\mathrm{jet}\,1,2}|<0.3;\quad0.3<|\eta^{\mathrm{jet}\,2,1}|<0.9$ & Forward-central \\  
 Topology C & $|\etajone|,|\etajtwo|<0.3$ & Central-central \\ 
 Topology D & $0.3<|\etajone|,|\etajtwo|<0.9\quad\etajone\cdot\etajtwo<0$ & Forward-backward.\\
 \hline
\end{tabular}
\caption{Regions in jet pseudo-rapidity employed in the STAR analysis of~\cite{STAR:2021mqa}. Here $\etajone$ and $\etajtwo$ refer, respectively, to the pseudo-rapidity of the two hardest jets in an event.  The notation $\eta^{\mathrm{jet}\,1,2}$ in the case of topology B indicates that \textit{either} the hardest \textit{or} the second hardest jet is measured in this region, with the other jet being measured in a different pseudo-rapidity range.
\label{tab:star-tops}
}
\end{center}
\end{table}
 
The different topologies have the advantage of probing the parton distribution functions of the two colliding protons at different values of the partons' momentum fractions $x_1$ and $x_2$.  For  the entire analysis jets are reconstructed with the anti-$k_T$ algorithm with a resolution parameter of $R=0.5$, and 
asymmetric selection cuts are applied on the transverse momenta of the two hardest jets, 
\beq
\label{eq:pt-jet-asymm}
\ptjone>7~\GeV, \qquad \ptjtwo>5~\GeV. 
\eeq
%

%
\begin{figure}[tp]
    \centering
    \includegraphics[width=\textwidth]{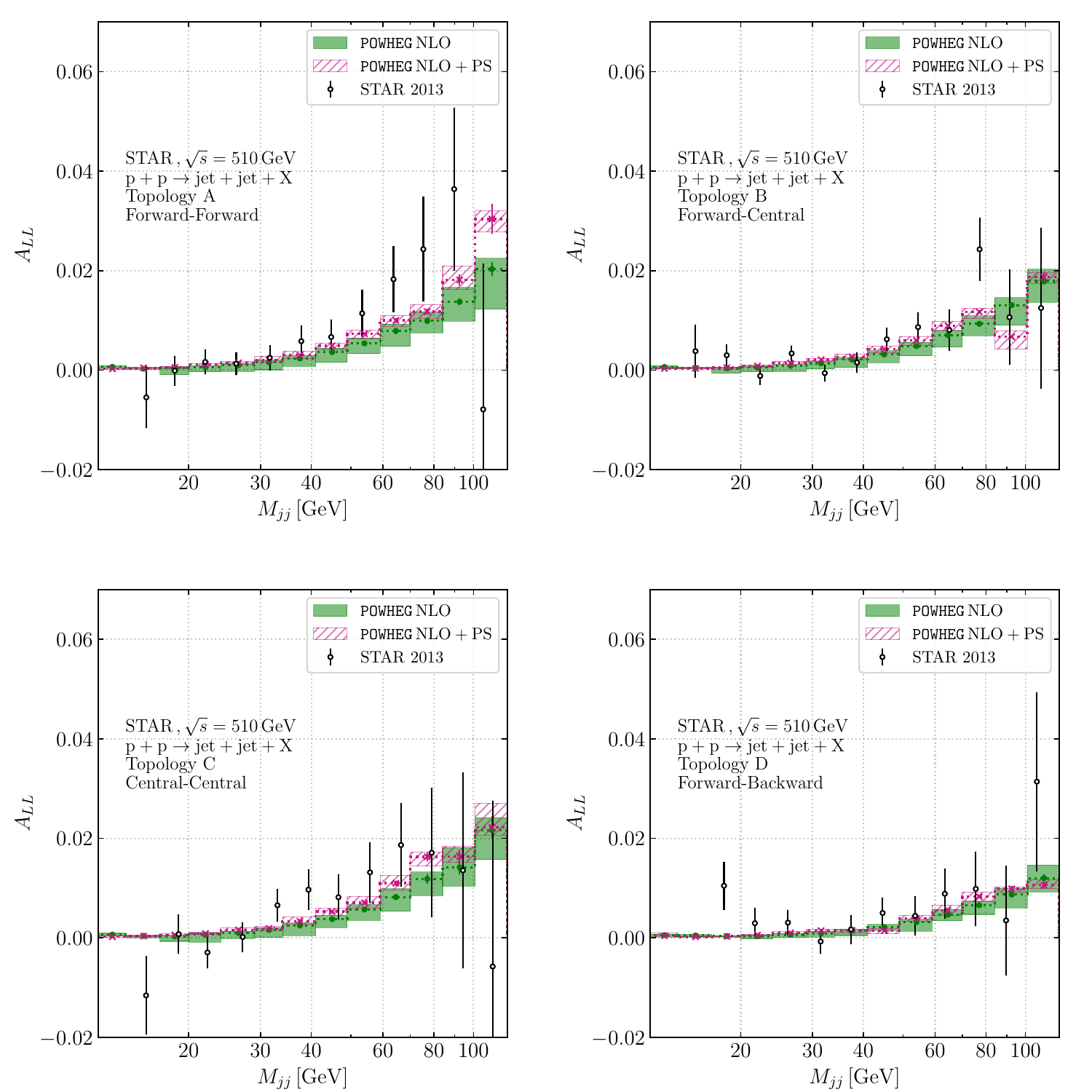}%
    \caption{
    \label{fig:STAR_ALL_510}
    Experimental results of the STAR collaboration~\cite{STAR:2021mqa} for the di-jet double-spin asymmetry at $\sqrt{s}=510~\mathrm{GeV}$ together with \PBOX{} predictions at NLO (green) and NLO+PS (dashed-pink).  Error bars are used to indicate the numerical uncertainty of the results while the bands indicate the theoretical uncertainty estimated by a 7-point scale-variation. 
    }
\end{figure}
%

In Fig.~\ref{fig:STAR_ALL_510} we compare our predictions to the di-jet measurements of the STAR collaboration at $\sqrt{s}=510~\mathrm{GeV}$. The plots present both the fixed-order NLO results  as well as their matching with a parton shower. As in the previous section and following the STAR analysis, the showered results are obtained using the default shower of \PYTHIAS{} and the Perugia 2012 tune, with the additional use of the standard hadronization feature and neglecting multi-parton interactions. In the case of the unpolarized cross section, needed to calculate the double spin asymmetry, the results were obtained using the same \PYTHIAS{} settings and the MSHT20 set of parton distribution functions~\cite{Bailey:2020ooq}. For our comparison to experimental results we omit presenting results at the {\tt LHE} level which, as mentioned in the previous section, can have a non-physical behavior for low values of the $R$ parameter.

Overall, the agreement of our predictions with data is already good at the NLO level, within the rather considerable experimental uncertainties. The effect of the parton shower is to some extent cancelled in the asymmetry, leading to a slight increase of the predicted double-spin asymmetry at larger values of the di-jet invariant mass. In the case of topologies A and C, this shift of approximately $25\%$ brings the theoretical prediction closer to the experimental measurements. It is worth noting that hadronization and multi-parton-interaction effects largely cancel in the ratio defining the spin asymmetry.

%
\begin{figure}[tp]
    \centering
    \includegraphics[width=\textwidth]{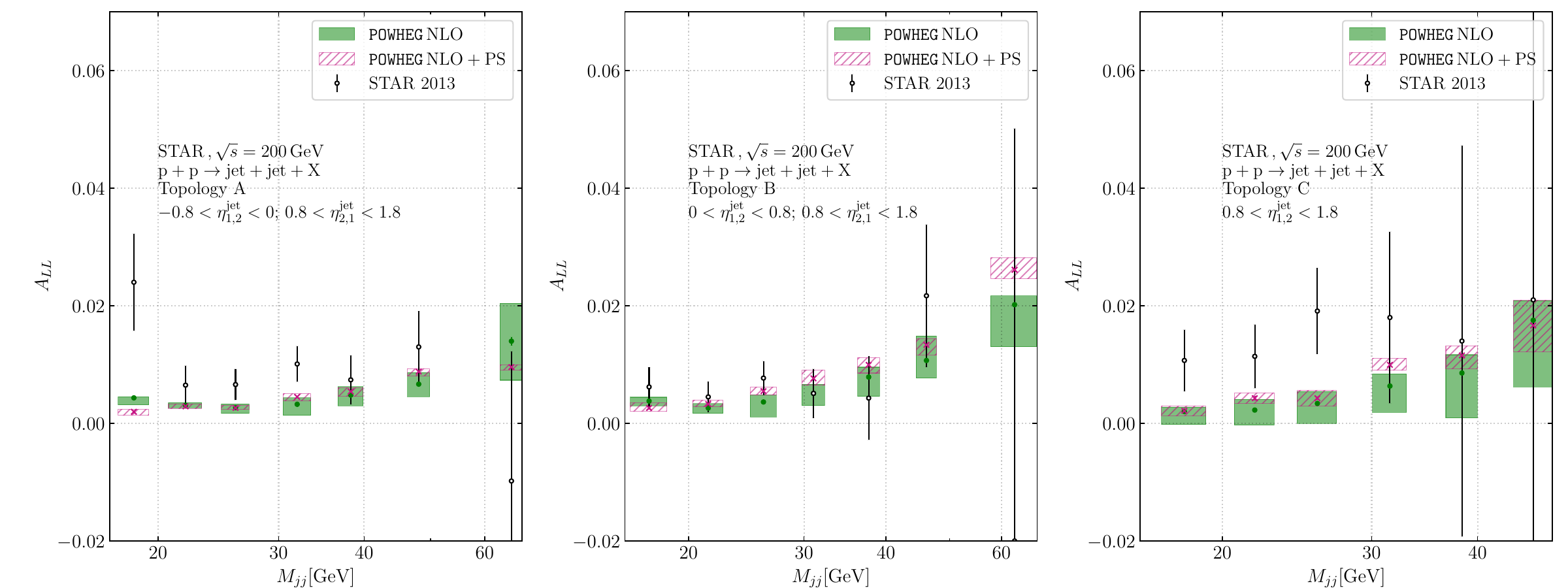}%
    \caption{
    \label{fig:STAR_ALL_200}
    Similar to Fig.~\ref{fig:STAR_ALL_510}, now for the STAR measurements~\cite{STAR:2025:DijetIntermediateALL:PRD112} at $\sqrt{s}=200~\mathrm{GeV}$.
    }
\end{figure}
%

A similar comparison for the STAR measurements at $\sqrt{s}=200~\mathrm{GeV}$, obtained during the 2015 run~\cite{STAR:2025:DijetIntermediateALL:PRD112}, is presented in Fig.~\ref{fig:STAR_ALL_200}. In this case, the data correspond to the di-jet double-spin asymmetries as a function of the invariant mass of the di-jet system in the pseudo-rapidity range $-0.8<\etaj<1.8$, and are divided into three distinct topologies depending on whether the jets are detected in the east half of the Barrel Electromagnetic Calorimeter ($-0.8<\etaj<0.0$), its west half ($0.0<\etaj<0.8$), or the Endcap ($0.8<\etaj<1.8$), according to \ref{tab:star-tops200}.

\begin{table}[t]
\begin{center}
\begin{tabular}{ | c | c |}
\hline
  $-0.8<\eta^{\mathrm{jet}\,1,2}<0.0;\quad0.8<\eta^{\mathrm{jet}\,2,1}<1.8$ & Topology A \\ 
  $0.0<\eta^{\mathrm{jet}\,1,2}<0.8;\quad0.8<\eta^{\mathrm{jet}\,2,1}<1.8$ & Topology B \\  
  $0.8<\etajone,\etajtwo<1.8$ & Topology C.\\
 \hline
\end{tabular}
\caption{Regions in jet pseudo-rapidity employed in the STAR analysis of~\cite{STAR:2025:DijetIntermediateALL:PRD112}. The notation follows the one of Tab.~\ref{tab:star-tops}.
\label{tab:star-tops200}}
\end{center}
\end{table}

Slightly different from the case of the data at $\sqrt{s}=510~\mathrm{GeV}$, the anti-$k_T$ algorithm here is employed with a resolution parameter of $R=0.6$, and asymmetric cuts are imposed on the transverse momenta of the jets, 
\beq
\label{eq:pt-jet-asymm200}
\ptjone>8~\GeV, \qquad \ptjtwo>6~\GeV. 
\eeq
The comparison shown in Fig.~\ref{fig:STAR_ALL_200} again demonstrates good agreement between theory and data at the NLO level. The inclusion of parton-shower effects leads to a mild enhancement of the asymmetry at larger invariant masses, while remaining  in general within the scale uncertainty. Overall, these results provide a consistent validation of the \PBOX{} framework for polarized di-jet observables at RHIC energies. 
\section{Conclusions and outlook}
\label{sec:conclusions}

In this work, we present a new Monte Carlo event generator for di-jet production in longitudinally polarized proton–proton collisions at NLO+PS accuracy, relevant to the RHIC spin program.  
Our event generator is implemented in the \POWHEGBOXVV, and based on the extension of the \POWHEGBOX{} framework to account for polarized initial-state particles from~\cite{Borsa:2024rmh}, thus extending previous work for the helicity-averaged case~\cite{Alioli:2010xa}. The program can be obtained from the \PBOX{} repository at 
%
\url{https://gitlab.com/POWHEG-BOX/V2/User-Processes/dijet_pol}

We have carefully validated the results of our new implementation through the comparison to previous fixed-order results as well as through an analysis of the first branching generated by \POWHEG{} and the effects of matching to a parton shower. We observed that, at the level of the cross-section and within the kinematical configurations relevant for RHIC, some exclusive distributions can show a pathological behavior at the fixed order, while \POWHEG{} yields physically meaningful results through the resummation of enhanced logarithmic contributions.

We have additionally performed phenomenological studies for the latest di-jet measurements reported by the STAR collaboration at c.m.s.~energies of $\sqrt{s}=200~\GeV$ and  $\sqrt{s}=510~\GeV$. While the effect of the parton shower is to some extent suppressed in the double spin asymmetries reported by STAR, we observe some improvement in the description of $\sqrt{s}=510~\GeV$ data at the NLO+PS level.

Our Monte-Carlo program therefore provides a useful tool to simulate di-jet production at longitudinally polarized hadron colliders. In particular it can support current and future analyses by the STAR and PHENIX collaborations at RHIC, offering a systematic alternative to LO-based reweighting approaches.

\section*{Acknowledgements}
We are grateful to Werner Vogelsang for valuable comments, to Daniel de Florian for providing the NLO code from \cite{deFlorian:1998qp} and to Carlo Oleari for helping with the \POWHEGBOX{}.
This work has been funded by the Deutsche
Forschungsgemeinschaft (DFG, German Research Foundation) through the Research Unit FOR 2926 (Project No. 409651613).

\bibliographystyle{JHEP}
\bibliography{dijets-pp-pol}

\end{document}